\documentclass{aa}
\usepackage[varg]{txfonts}
\usepackage{hyperref}
\usepackage{amsmath}
\usepackage{relsize}
\usepackage{dsfont}
\usepackage{subcaption}

\begin{document}
\title{The spatial evolution of young massive clusters}
\subtitle{II. Looking for imprints of star formation in NGC\,2264 with Gaia DR2}

\author{Anne S.M. Buckner\inst{1}, Zeinab Khorrami\inst{2}, Marta Gonz\'alez\inst{3}, Stuart L. Lumsden\inst{1}, Estelle Moraux\inst{3}, Ren\'e D. Oudmaijer\inst{1}, Paul Clark \inst{2}, Isabelle Joncour\inst{3,5}, Jos\'e Manuel Blanco\inst{4}, Ignacio de la Calle\inst{4}, \'Alvaro Hacar\inst{6}, Jos\'e M. Herrera-Fernandez\inst{4}, Fr\'ed\'erique Motte\inst{3}, Jes\'us Salgado\inst{4} and Luis Valero-Mart\'in\inst{4}}

\institute{School of Physics and Astronomy, University of Leeds, Leeds LS2 9JT, U.K. \\ \email{a.s.m.buckner@leeds.ac.uk}
\and School of Physics and Astronomy, Cardiff University, The Parade, CF24 3AA, U.K. \and Universit\'e Grenoble Alpes, CNRS, IPAG, 38000 Grenoble, France \and Quasar Science Resources, S.L., Edificio Ceudas, Ctra. de La Coru\~na, Km 22.300, 28232, Las Rozas de Madrid, Madrid, Spain \and Department of Astronomy, University of Maryland, College Park, MD 20742, USA \and Leiden Observatory, Niels Bohrweg 2, 2333 CA Leiden, The Netherlands}

\date{Received / Accepted}


\abstract{Better understanding of star formation in clusters with high-mass stars requires rigorous dynamical and spatial analyses of star-forming regions.}{We seek to demonstrate that `INDICATE' is a powerful spatial analysis tool which when combined with kinematic data from Gaia DR2 can be used to probe star formation history in a robust way.}{We compared the dynamic and spatial distributions of young stellar objects (YSOs) at various evolutionary stages in NGC\,2264 using Gaia DR2 proper motion data and INDICATE.} {The dynamic and spatial behaviours of YSOs at different evolutionary stages are distinct. Dynamically, Class II YSOs predominately have non-random trajectories that are consistent with known substructures, whereas Class III YSOs have random trajectories with no clear expansion or contraction patterns. Spatially, there is a correlation between the evolutionary stage and source concentration: 69.4$\%$ of Class 0/I, 27.9$\%$ of Class II, and 7.7$\%$ of Class III objects are found to be clustered. The proportion of YSOs clustered with objects of the same class also follows this trend. Class 0/I objects are both found to be more tightly clustered with the general populous/objects of the same class than Class IIs and IIIs by a factor of 1.2/4.1 and 1.9/6.6, respectively. An exception to these findings is within $0.05^{o}$ of S\,Mon where Class III objects mimic the behaviours of Class II sources across the wider cluster region. Our results suggest (i) current YSOs distributions are a result of dynamical evolution, (ii) prolonged star formation has been occurring sequentially, and (iii) stellar feedback from S\,Mon is causing YSOs to appear as more evolved sources.} {Designed to provide a quantitative measure of clustering behaviours, INDICATE is a powerful tool with which to perform rigorous spatial analyses. Our findings are consistent with what is known about NGC\,2264, effectively demonstrating that when combined with kinematic data from Gaia DR2 INDICATE can be used to study the star formation history of a cluster in a robust way.}


\authorrunning{Buckner et al.}

\maketitle

\section{Introduction}\label{sect_intro}

With the second instalment of the Gaia survey (DR2; \citealt{2018A&A...616A...1G}), high-precision position and kinematic data became available for a large number of young clusters which previously lacked reliable parallax and proper motion measurements. Now, it is possible to probe the dynamical evolution and star formation history of these clusters through substructure, mass segregation, and relative dynamics studies of their young populations.

One of the fundamental questions in such analyses is to what degree stars `cluster' together and how does this change as the cluster evolves. The anwser to this question requires a combined study of the spatial intensity, correlation, and distribution of stars/clumps with the kinematic data. For this type of characterisation the use of local indicators \citep{Anselin1995LISApaper} is suggested. Unlike global indicators (e.g. the two-point correlation function) that derive a single parameter for a group of stars as a whole, local indicators derive a parameter for each unique source such that variations and trends as a function of fundamental parameters (stellar mass, evolutionary stage, position, and individual dynamical histories) can be distinguished. Unfortunately local indicators have remained largely ignored in cluster analysis due to a distinct lack of appropriate astro-statistics tools, and the best understood methods from other fields cannot be easily applied to (or are simply invalid for) astronomical datasets.

Our aim with this paper series is the development and application of local statistic tools, optimised for stellar cluster analysis. In Paper I \citep{2019A&A...622A.184B} we introduced the tool INDICATE (INdex to Define Inherent Clustering And TEndencies) to assess and quantify the degree of spatial clustering of each object in a dataset, demonstrating its effectiveness as a tracer of morphological stellar features in the Carina Nebula (NGC\,3372) using positional data alone. In this paper we demonstrate that when combined with kinematic data from Gaia DR2, INDICATE is a powerful tool to analyse the star formation history of a cluster in a robust manner.

Embedded in the Mon OB1 cloud complex, NGC\,2264 is located at a Gaia DR2 determined distance of $723^{+56}_{-49}$pc \citep{2018A&A...618A..93C}. Structurally, the cluster is elongated along a NW-SE orientation with two sub-clusters C and D in the southern region (Figure \ref{fig_images}). There is an age spread of $\sim$3-4\,Myr between the older star formation inactive northern region which contains the bright O-type binary star, S\,Mon, and the younger ongoing star formation southern region within the C and D sub-clusters (\citealt{2008MNRAS.386..261M}, \citealt{2017MmSAI..88..848V}). Recent studies have found NGC\,2264 to be rich in YSOs of all evolutionary stages (\citet{2012A&A...540A..83T}, \citet{2013ApJS..209...31P}, \citet{2014ApJ...794..124R}, \citealt{2018A&A...609A..10V}). Moreover, owing to its close proximity, this cluster is one of the best researched in the literature with numerous studies into its recent and ongoing star formation (for example \citealt{2009AJ....138.1116S}, \citealt{2010AJ....140.2070S}, \citealt{2012A&A...540A..83T}, \citealt{2017MmSAI..88..848V}, \citealt{2017MNRAS.465.1889G}, \citealt{2018A&A...609A..10V}). As such we have chosen to focus our efforts on NGC\,2264 as (i) the validity of our results can be checked against what is already known about the cluster, and (ii) its large YSO population makes this cluster an ideal candidate to show that INDICATE can successfully provide the rigorous spatial analysis necessary to validate and correctly interpret dynamical behaviours found with DR2 data for young clusters.  

The paper is structured as follows. We introduce our sample of YSOs in Section\,\ref{sect_sample} and analysis methods in Section\,\ref{sect_method}. In Section\,\ref{sect_results} we present the results of our spatial and kinematic analyses, which we discuss in Section\,\ref{sect_discuss}. A conclusion is given in Section\,\ref{sect_summary}.

\section{Sample of young stellar objects}\label{sect_sample}

In this section we describe our sample of young NGC\,2264 members. The following terminology is employed to discriminate between the different evolutionary stages of YSOs:

\begin{itemize}

\item Class 0: protostars without dust emission
\item Class I: protostars with envelope and disc dust emission
\item Class II: Pre-main sequence (PMS) stars with circumstellar accretion discs
\item Class TD: transition discs; an intermediate stage between Class II and III where the disc has a radial gap
\item Class III: PMS stars without discs

\end{itemize}

\subsection{Catalogue selections}\label{sect_cat}

\begin{figure*}
\centering
   \includegraphics[height=0.45\textheight]{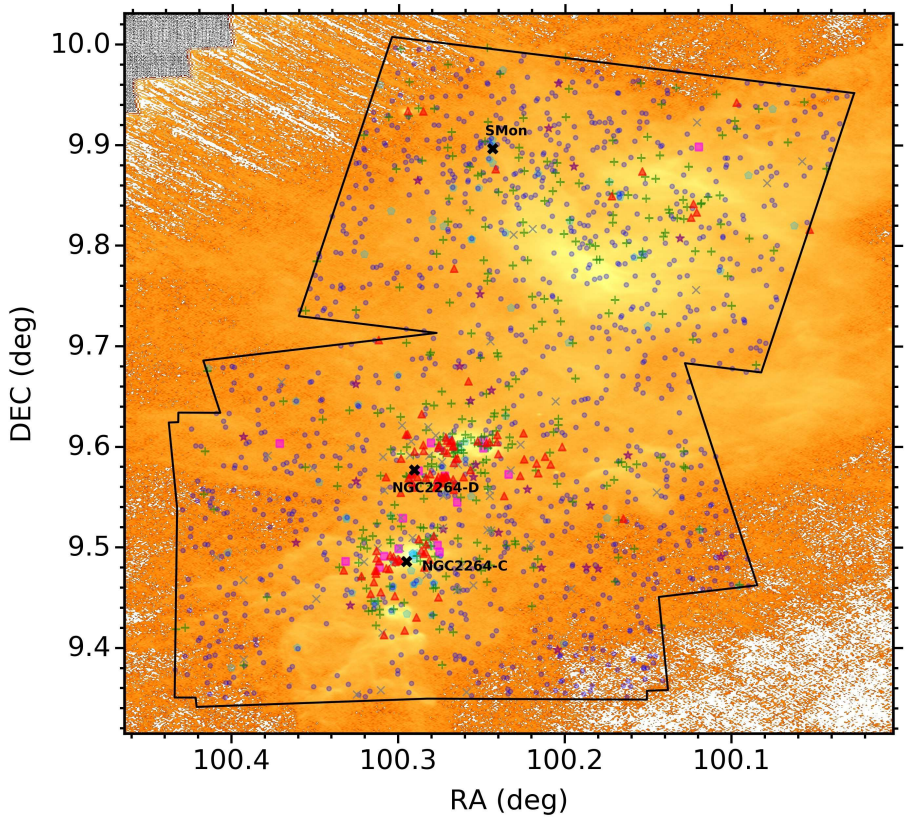} 
  \caption{Plot showing the distribution of our sample overlaid on the Herschel 70$\mu$m map of the region to clearly show the locations of S\,Mon and sub-clusters C, D. Colours/shapes: Class 0/I (red triangles), Class II (green plus signs), Class TD (purple stars), Class III/F (dark blue dots), Class II/III (grey crosses), Class Ambiguous (pink squares), and non-YSO (light blue pentagons).}  \label{fig_map} 
\end{figure*}

We draw our sample from two independent catalogues of the region. The first was constructed by \citet{kuhn_spatial_2014} (hereafter K14) as part of the MYStIX project \citep{feigelson_overview_2013} which surveyed 20 OB-dominated young clusters using a combination of Spitzer IRAC \citep{2004ApJS..154...10F} infrared and Chandra \citep{2000SPIE.4012....2W} X-ray photometry. The bounds of the $\sim\,0.19^{\circ}$ NGC\,2264 region  surveyed (Figure \ref{fig_map}) correspond to the limits of the Chandra mosaic coverage, as this is smaller than that of the Spitzer IRAC photometry.  Identification and classification of YSOs were carried out by \citet{2013ApJS..209...31P} who used Spitzer IRAC, 2MASS \citep{2006AJ....131.1163S}, and UKIRT \citep{2007MNRAS.379.1599L} imaging photometry with spectral energy distribution fitting to flag sources as `0/I', `II/III', `non-YSO (stellar)' or `Ambiguous (YSO)'. The catalogue consists of 969 sources of which 139 are Class 0/I, 298 Class II/III, 413 non-YSO (stellar), and 119 Ambiguous (YSO).

The second catalogue is by \citet{2014ApJ...794..124R} (hereafter R14) who analysed 2MASS and Spitzer photometry of Mon OB1 East to identify YSO members. A three-phase classification method by \citet{2009ApJS..184...18G} which utilised photometry in eight infrared bands ($J$, $H$, $K$, $3.6\,\mu$m, $4.5\,\mu$m, $5.8\,\mu$m, $8.0\,\mu$m, and $24\,\mu$m) was employed, resulting in 10454 potential candidates. We selected only sources in the smaller K14 region of which there are 1645 comprised of 70 Class 0/I, 307 Class II, 26 Class TD, 1189 Class III/F (where `F' denotes source is potentially a line-of-sight field star: see next section), and 53 contaminants (AGN, Shock, PAH). 

We merged the two samples and performed a cross-match to remove duplicates, finding 1848 unique sources for the region. There is significant overlap between the two catalogues with a R14 counterpart for 766/969 K14 sources and a discrepancy in the assigned classifications of $24.4\%$ (Table\,\ref{appen_duplic}). Interestingly, 101/119 of the duplicate sources flagged as Ambiguous (YSO) in K14 have a definitive classification from R14 (i.e. 0/I, II, TD, III/F, AGN, Shock, PAH). Sources classified by R14 form the majority of the merged sample so we adopted their classifications for all sources that appear in both catalogues. This is statistically justified as we are interested in the spatial behaviour of the YSO population as a whole, not individual sources. Thus (i) a single classification system should be utilised, where possible, to ensure the spatial analyses are systematic; and (ii) if an incorrect classification is assumed for an individual source, it would effectively be an outlier so does not have a statistically significant impact on our results as our methodology is robust against outliers (Sect.\,\ref{sect_indicate}). 

After removing the contaminants, our final sample contains 1795 sources. Table\,\ref{table_gaia} details its composition and Figure\,\ref{fig_map} its distribution. We create four sub-samples: $S_1$ - all sources ($n_1=1795$); $S_2$ - Class 0/I only ($n_2=111$); $S_3$ - Class II only ($n_3=307$); $S_4$ - Class III/F only ($n_4=1189$). 

\subsection{Field star contamination}\label{sect_contam}

The classification method employed by R14 distinguishes Class III sources from earlier type YSOs by their (lack of) 3.6$\mu$m and 4.5$\mu$m excess emission. However, as field stars in the line of sight also lack this excess they cannot be readily distinguished from true Class III cluster members. To estimate the number of field star contaminants, the authors calculated the expected number of field stars from comparison to two control regions neighbouring Mon OB1 East, determining a contamination of $\sim 29\%$ in the region of NGC\,2264. Thus 345 of the 1189 Class III/F sources in our sample are expected to be field stars and 844 Class III members.  

For an independent measure, we cross-matched our sample with the Coordinated Synoptic Investigation of NGC\,2264 (CSI\,2264; \citealt{2014AJ....147...82C}). Uniquely, CSI\,2264 continuously observed the region for 30 days using Spitzer IRAC and the Convection, Rotation and Planetary Transits satellite (CoRoT; \citealt{2006cosp...36.3749B}) simultaneously, with additional observations from 13 other telescopes. Subsequently the CSI\,2264 photometric database is one of the most comprehensive of the region to date containing an impressive 146,\,855 sources. Membership of each source was assessed against photometric, spectroscopic, spatial, and kinematic criteria (see Appendix A.1 of \citealt{2014AJ....147...82C} for details) and flagged as `very likely member', `possible member', `likely field object' or `no membership information'. For our 1189 Class III/F objects, 775 are very likely or possible members, 91 are likely field objects, 320 have no membership information and 3 do not appear in the catalogue. Of the 320 with no membership information, 11 are identified by K14 as members of the cluster indicating a field contamination rate of $8-34\%$. This is in good agreement with the estimate of R14 of $29\%$ which suggests the true contamination is towards the upper limit of this range. We therefore conclude the R14 contamination calculation is reasonable and assume this value for our analysis (but see Sect\,\ref{set_stat_com}).

\subsection{Completeness of sample}\label{sect_complete}

\begin{figure*}
\centering
   \includegraphics[height=0.32\textheight]{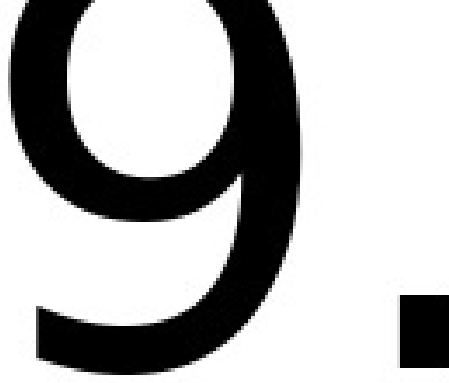}\hfill
   \includegraphics[height=0.32\textheight]{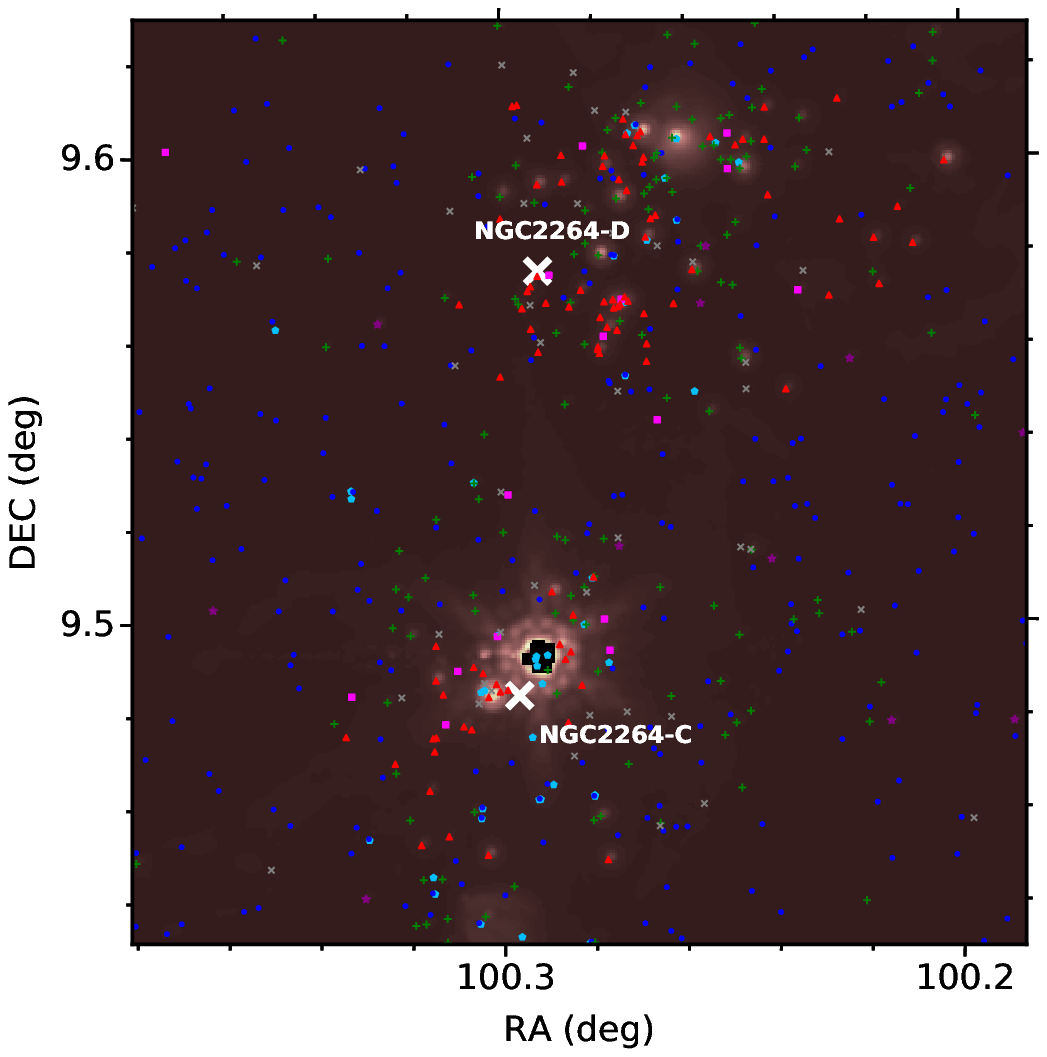}
  \caption{(Left panel) Sample overlaid on a Herschel $250\,\mu$m image of NGC\,2264. White lines and crosses denote the borders and sources of the sample respectively. (Right panel) Zoomed-in plot of Figure\,\ref{fig_map} in the NGC\,2264-C and -D sub-cluster region, overlaid on a corresponding Spitzer $24\,\mu$m image. Colours and symbols as defined in Figure\,\ref{fig_map}.}  \label{fig_images}
\end{figure*}

We anticipate two sources of incompleteness in our sample. The first relates to the heavy extinction present in the cluster owing to its embedded nature, which can be seen in the Herschel $250\,\mu$m image of NGC\,2264 shown in the left panel of Figure \ref{fig_images}. Despite compiling our sample from two catalogues of the region, in the absence of longer wavelength photometry it is reasonable to assume that they suffer from incompleteness due to dust extinction. In particular, we anticipate the majority of ‘missing’ sources are located in regions of highest extinction (sub-clusters C and D) and for these to primarily be the most deeply embedded Class 0/I objects which have not been detected. Indeed, Class 0/I objects constitute only $16\%$ of sources in the right panel of Figure \ref{fig_images}. To gauge how many Class 0/I sources are `missing' we consult submillimetre surveys of the sub-clusters (\citealt{2006A&A...445..979P}, \citealt{2007ApJ...667L.179T}), and find there are at least 16 Class 0/I sources which are not included in our sample. We refrain from appending our sample to ensure it remains homogeneous (and thus results reliable) as these surveys only cover relatively small areas of NGC\,2264's southern region. However it should be noted that the inclusion of these highly concentrated sources would strengthen, rather than diminish, the trends found in Sect.\,\ref{sect_results} and thus our conclusions remain unchanged irrespective of our decision to exclude these objects.

The second relates to the point spread function (PSF) wings of a bright star at the centre of NGC\,2264-C, as seen in the Spitzer MIPS $24\,\mu$m image (Figure \ref{fig_images} right panel). There is significant angular dispersion of the PSF which likely occludes a number of fainter sources in 2MASS and Spitzer IRAC bands (from which both catalogues were derived).

\subsection{Gaia DR2 kinematic data}\label{sect_dr2}

We cross-match all 1795 sources with the Gaia DR2 database by colour and position, finding proper motions for 1268 sources and radial velocities for 20 sources (Table\,\ref{table_gaia}). As expected there is a distinct lack of Class 0/I objects with kinematic data in DR2 because the magnitudes of these deeply embedded objects are typically being below the Gaia detection limit. Before proceeding we must consider the impact of systematic errors in the proper motion measurements caused by the Gaia scanning law as our sample occupies an area $<<1^{\text{o}}$ and spatial scales $<1^{\text{o}}$ are most affected with an root mean square amplitude of 0.066 mas yr$^{-1}$ (\citealt{2018A&A...616A...2L}). To ensure these errors do not dominate the measurements it is necessary to exclude any sources from our kinematic analysis in Sect.\,\ref{sect_results_gaia} with a proper motion (within error bounds) of $<$0.066 mas yr$^{-1}$. A search of the sample reveals 29 sources that meet the exclusion criteria. We further exclude 672 objects with $r_{hi}<674$pc or $r_{lo}>779$pc, where $r_{hi}$ is the upper error bound and $r_{lo}$ the lower error bound on the distance estimate determined by \citet{2018AJ....156...58B} from DR2 parallaxes. The cut-off distance values correspond to the upper/lower DR2 distance estimate of $723^{+56}_{-49}$pc for NGC\,2264 \citep{2018A&A...618A..93C}.

Two perspective corrections on the proper motions are needed prior to analysis: first, the radial motions of members cause NGC\,2264 to appear to contract and, second, members appear to move towards a point of common convergence as they are part of the same stellar system and share a common motion. The former is corrected using Eq.13 of \citet{vanLeeuwen_2009} and the latter by subtracting the mean proper motion of the system from observed proper motions of each member. Appendix\,\ref{append_gaia_pm} describes these calculations. 

Distance measurements from Gaia DR2 suggest a significant proportion (561/966) of Class III objects are not true members, which is considerably higher than the $29\%$ identified from photometric analysis (R14). While this may reflect the number of true members of the cluster it may also be a symptom of a number of observational biases (e.g. imprecise/too strict distance criteria, unresolved binaries etc.).  In addition approximately one-third of Class III objects in our sample lack parallax measurements from which to make a distance-dependant membership determination. As such, it is important to ascertain the effect of a significantly reduced Class III sample size on our results; for example whether the spatial trends found in the Sect.\,\ref{sect_results_s1} and \ref{sect_results_s2} hold for these objects identified by the kinematic data. To check, we re-ran our spatial analyses (as outlined in Sect.\,\ref{sect_results}) excluding all Class III sources that did not meet our above discussed DR2 distance criteria and find our conclusions on the spatial behaviour of YSOs in NGC\,2264 are unaffected by the exclusion of sources that fail the distance criteria. 

Therefore as only $79.0\%$ of Class III, $74.9\%$ of Class II, and $1.8\%$ of Class 0/I sources have reliable DR2 data we used the full sample with Monte Carlo sampling described in Section \,\ref{set_stat_com} for our spatial analyses. For our kinematic analysis we only used sources which met our DR2 distance criteria (Table\,\ref{table_gaia}) to ensure the proper motion patterns we observe for Class II and III objects are an accurate reflection of typical member motions.

\begin{table}
\caption{Number of sources in our sample by Class, with Gaia DR2 proper motion (PM) and radial velocity (RV) measurements and that have been excluded/included from our kinematic analysis in Section\,\ref{sect_results_gaia}.}              
\label{table_gaia}      
\centering                                      
\begin{tabular}{c c c c c c}          
\hline\hline   
Class & Total & PM & RV & Excluded & Included \\
\hline
0/I & 111 & 2 & 0 & 1 & 1\\
II& 307 & 232 & 1 & 85 & 147\\
TD & 26 & 23 & 0 & 5 & 18 \\
III/F & 1189 & 966 & 19 & 588 & 378\\
II/III & 60 & 19 & 0 & 10 & 9\\
Ambiguous(YSO)& 17 & 1 & 0 & 1 & 0\\
non-YSO (stellar)& 85 & 25 & 0 & 11 & 14\\
\hline 
& 1795 & 1268 & 20 & 701 & 567\\                     
\hline
\end{tabular}
\end{table}

\section{Analysis method}\label{sect_method}

\subsection{INDICATE}\label{sect_indicate}

We analyse the spatial distributions of YSOs using INDICATE \citep{2019A&A...622A.184B}. A statistical clustering too, INDICATE is used to study the intensity, correlation, and spatial distribution of point processes in 2+D discrete astronomical datasets.  It is a local statistic which quantifies the degree of association of each point in a dataset through a comparison to an evenly spaced control field. Advantageously, INDICATE does not make assumptions about (or require a priori knowledge of) the shape of the distribution, nor the presence of any substructure. Extensive statistical testing has shown it to be robust against outliers and edge effects, and independent of the size and number density of a distribution \citep{2019A&A...622A.184B}. 

When applied to a dataset of size \textit{n}, INDICATE derives an index for each data point $j$, defined as

	\begin{equation} \label{eq_I}
	\\ I_{5,\,j}= \frac{N_{\bar{r}}}{5}
	\end{equation}

where $N_{\bar{r}}$ is the number of nearest neighbours to data point $j$ within a radius of the mean Euclidean distance, $\bar{r}$, of every data point to its $5^{\text{th}}$ nearest neighbour in the control field. The index is a unit-less ratio with a value in the range $0 \le I_{5,\,j} \le \frac{\textit{n}-1}{5}$ such that the higher the value, the more spatially clustered a data point. For each dataset the index is calibrated so that significant values can be identified. To do this, 100 realisations of a random distribution of the same size \textit{n}, and in the same parameter space, as the dataset are generated. Then INDICATE is applied to the random samples to identify the mean index values of randomly distributed data points, $\bar{I_5}^{random}$. Point $j$ is considered spatially clustered if it has an index value above a `significance threshold', $I_{sig}$, of three standard deviations, $\sigma$, greater than $\bar{I_5}^{random}$ i.e.

	\begin{equation}\label{eq_IjIsig}
	\\ I_{5,\,j}> I_{sig} \text{\,,\,\,\,\,\,where\,\,\,\,\,} I_{sig}=\bar{I_5}^{random}+3\sigma
	\end{equation}

Table\,\ref{table_Isig} lists the significance thresholds for $S_{1}$ to $S_{4}$. Appendix\,A of \citet{2019A&A...622A.184B} gives an in-depth discussion of the behaviour and properties of the index in random distributions.

\subsection{Statistical considerations of field star contamination}\label{set_stat_com}

As discussed in Section\,\ref{sect_contam}, $29\%$ (345) of the 1189 Class III/F sources in our sample are expected to be field stars. To ensure our spatial analysis results are reflective of the behaviour of Class III YSOs we randomly remove 345 sources flagged as `III/F’ from the $S_1$ and $S_4$ samples prior to analysis. We limited sources that we removed to those which have not also been identified by K14 as YSO (0/I, II/III, Ambiguous). After analysis the sources are replaced, and the process repeated for a total of 100 iterations. Statistics presented in Section\,\ref{sect_results} for $S_1$ and $S_4$ are representative of mean values derived over the 100 samples. The significance thresholds given for $S_1$ and $S_4$ in Table\,\ref{table_Isig} were determined for sample sizes of 1450 and 844 respectively. The maximum difference of mean index values for each iteration is $\bar{I_{5}}<0.1$ for both samples. We find that changing the contamination rate to the higher estimate of $34\%$ (Sect.\,\ref{sect_contam}) has a negligible impact; the trends found in Sect.\,\ref{sect_results_s1} and \ref{sect_results_s2} are unchanged and the difference between mean index values for the two rates is $\bar{I_{5}}<0.1$.

\begin{table}
\caption{Significance threshold, $I_{sig}$, of index values for objects in each sample determined using Eq.\,\ref{eq_IjIsig}. Above this value an object in the sample is considered spatially clustered.}
\label{table_Isig}      
\centering                                      
\begin{tabular}{c c c c c}        
\hline\hline                       
& $S_{1}$ & $S_{2}$ & $S_{3}$ & $S_{4}$ \\ 
\hline 
$I_{sig}$ & 2.3 & 2.2 & 2.3 & 2.3 \\
\hline 

\end{tabular}
\end{table}

\section{Results}\label{sect_results}

\subsection{Distribution of the YSO population}\label{sect_results_s1}

We applied INDICATE to the $S_1$ sample to investigate the clustering behaviour of YSOs in NGC\,2264. As expected the majority of clustered stars are located in the southern region within the star formation active NGC\,2264-C and D sub-clusters, whereas clustering in the older northern region is primarily found in the vicinity of S\,Mon. There is a distinct relationship between evolutionary stage and clustering behaviour of the YSOs. The number of Class 0/I objects with an index above the significance threshold in $S_{1}$ is $69.4\%$, in contrast to $27.9\%$ for Class II, $11.5\%$  for Class TD, and $7.7\%$ of Class III members. Furthermore, there is also a relationship between the degree to which YSOs are clustered (number of neighbours in local neighbourhood) and class; spatially clustered YSOs have median $I_{5}$ values of 5.2 for Class 0/I, 4.2 for Class II, 3.2 for Class TD, and 2.8 for Class III. This implies that first, the more evolved an object is the less likely it is to be clustered and second, more evolved objects that are clustered are less concentrated and more dispersed than their less evolved counterparts. 

\begin{figure}
\centering
   \includegraphics[width=0.5\textwidth]{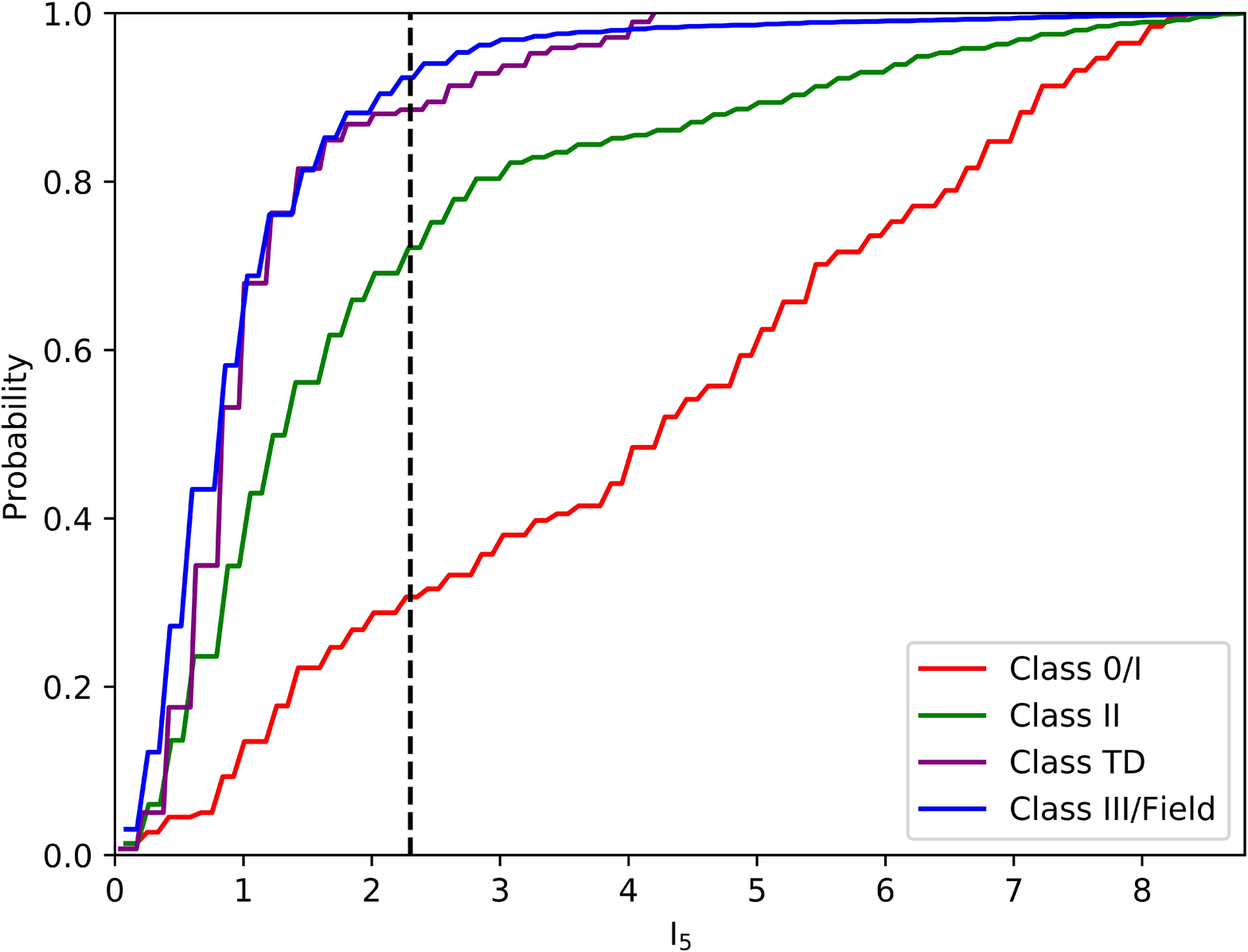}
  \caption{Empirical Cumulative Distribution Function (ECDF) of index values, $I_{5}$, calculated for sample $S_1$. The dashed black line denotes the significance threshold of the sample (Table\,\ref{table_Isig}). The intercept between the significance threshold and ECDFs is equal to 1-F, where F is the fraction of sources YSOs with an index value greater than this ($I_{5}>I_{sig}$) for each class. As can be seen, the ECDFs of each class are distinct which indicates the differences in their clustering behaviours are significant.}   \label{fig_cdf}
\end{figure}

To measure whether these trends are real and significant we compare the index values derived for the different classes using two sample Kolmogorov–Smirnov Tests (2sKSTs) with a strict significance boundary of $p < 0.01$. The null hypothesis of this test is that differences in the comparative clustering behaviours of two classes are not significant, so their index values have similar empirical cumulative distribution functions (ECDFs). The similarity is quantified by the 2sKST statistic, $D$, as the distance between two ECDFs (the smaller the statistic, the more similar the distributions). Figure \ref{fig_cdf} shows the ECDFs of Class 0/I, II, TD, and III objects in $S_1$ to be dissimilar, and this is confirmed by the 2sKSTs ($p << 0.01$). We therefore reject the null hypothesis: Class 0/I, II, TD, and III objects have distinct clustering behaviours and our finding that clustering behaviour is a true function of evolutionary stage is both real and significant.

Our assertion is further strengthened by the ECDFs of Class TD and III objects, which are distinct but closely resemble each other ($D_{\text{\,TD}, \text{\,III}}=0.1$). As Class TD objects represent an intermediate evolutionary stage from Class II to III it is reasonable to expect these objects to demonstrate the most similar clustering traits to the Class III objects (the next evolutionary stage).

\subsection{Spatial behaviour within classes} \label{sect_results_s2}

We now apply INDICATE to the $S_2$, $S_3$, and $S_4$ samples to evaluate the tendency for objects of the same class to cluster together. Table \ref{table_stats} summarises the statistics of the index values derived for each sample. There is a distinct trend between class and proportion of objects with an index above the significance threshold: $84.7\%$ (Class 0/I), $35.2\%$ (Class II), $2.8\%$ (Class III). In addition, Class 0/I objects are also found to be typically more tightly clustered together than Class II and III objects by a factor of 4.1 and 6.6 respectively. The maximum number of nearest neighbours of the same class decreases with increasing evolutionary stage from 61 (Class 0/I) to just 20 (Class III). This implies that first, the less evolved an object is the more likely it is to be clustered with objects of the same class and second, less evolved objects that are spatially clustered are typically more tightly concentrated together and less dispersed than their more evolved counterparts.

An exception to this trend is in the vicinity of the northern O-type binary, S\,Mon. In this region we find Class III objects in the local neighbourhood of S\,Mon to be significantly more self-clustered than the wider NGC\,2264 region and exhibit spatial behaviour patterns comparable to that of Class II objects. Sources have higher values than typical within a radius of $0.1^{o}$ of S\,Mon, but in particular within $0.05^{o}$ which contains the most spatially clustered Class III objects in the whole sample. Here $29.1\%$ of objects have an index above the significance threshold, with the median index value of those objects being $\tilde{I_5}=2.0$, which is comparable with Class II's across the NGC\,2264 region (Table \ref{table_stats}).

\subsection{Kinematic behaviour within classes} \label{sect_results_gaia}

We examine the magnitudes of proper motion for our distance selected sample (Section\,\ref{sect_dr2}), finding it to have median value of 1.131 mas yr$^{-1}$ with 1.009 mas yr$^{-1}$ and 1.192 mas yr$^{-1}$ for Class II and III sources respectively. 

\begin{figure*}
\centering
   \includegraphics[height=0.45\textheight]{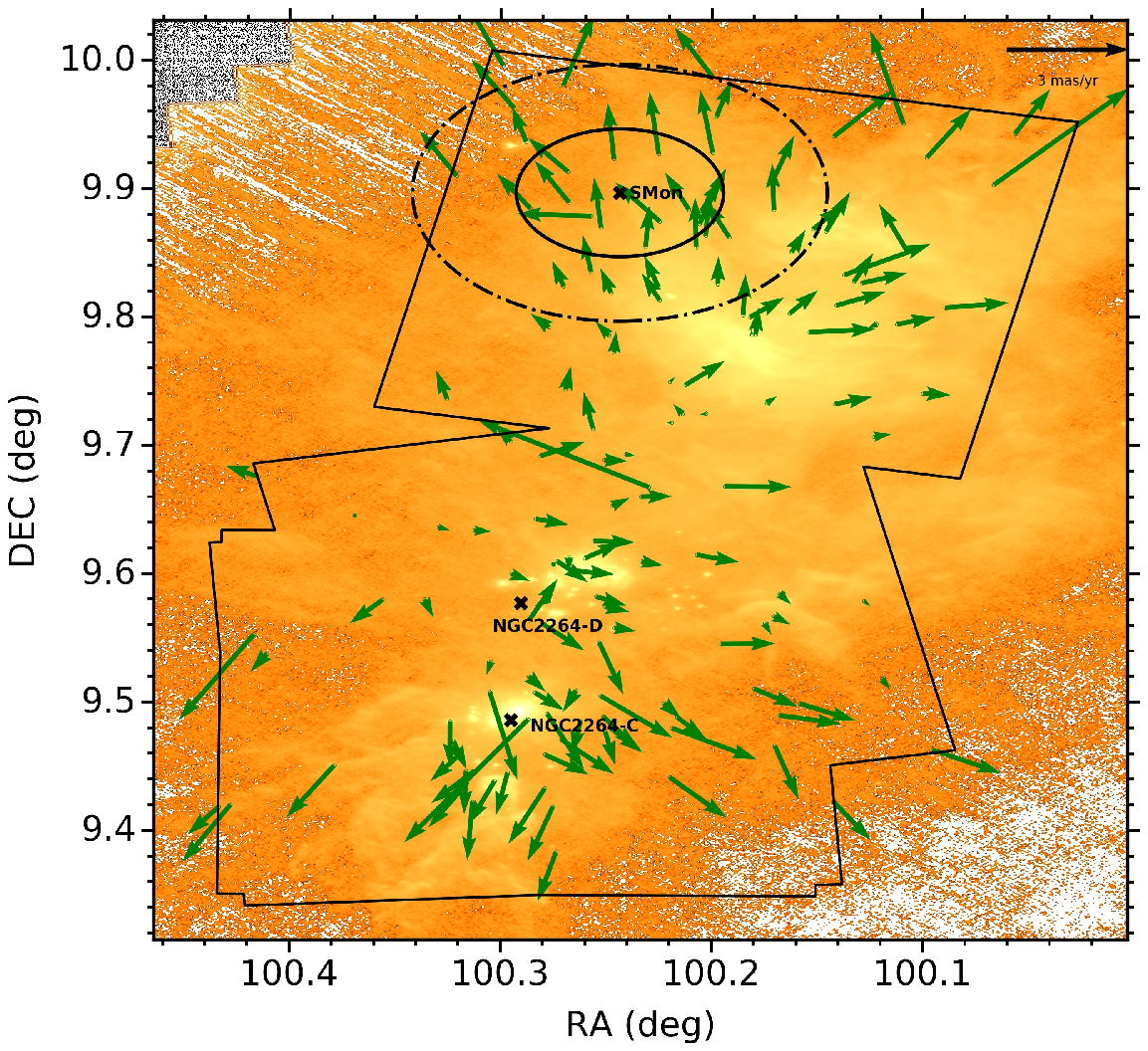} \hfill
   \includegraphics[height=0.45\textheight]{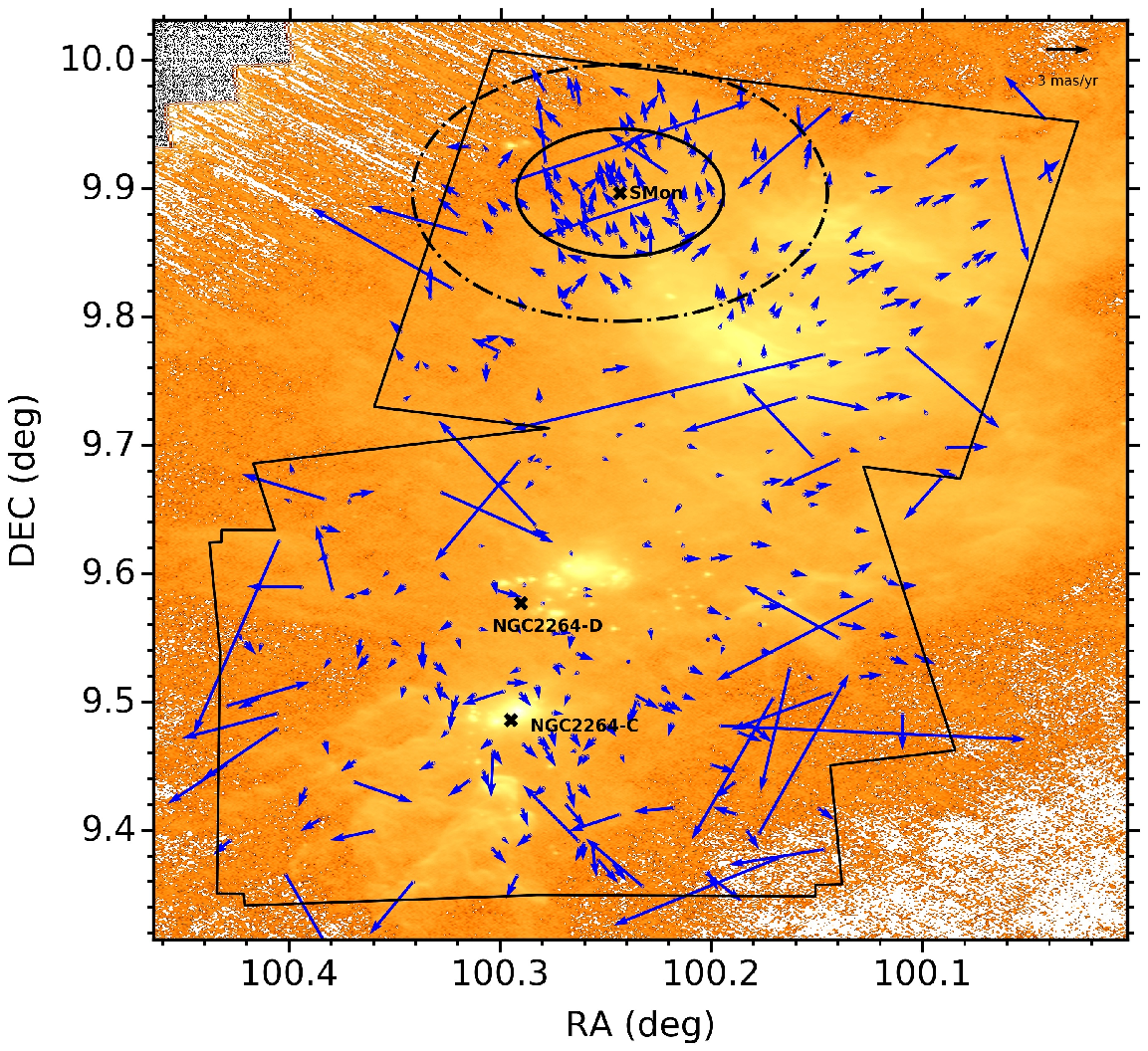} \hfill
  \caption{Distribution of Gaia proper motions for (\textit{top panel}) Class II and (\textit{bottom panel}) Class III objects of our sample  overlaid on the Herschel 70$\mu$m map of the region to clearly show the locations of S\,Mon and sub-clusters C, D for the reader's reference (outliers not shown). The black solid and dot-dash ellipses around S\,Mon represent radii of $0.05^{o}$ and $0.1^{o}$ respectively.}  \label{fig_pm} 
\end{figure*}

The kinematic distribution of Class II and III sources shown in Figure\,\ref{fig_pm} are consistent with our findings in Sect.\,\ref{sect_results_s1}. Motions of Class III sources are dispersed and randomised with no clear expansion or contraction patterns in the southern region. In the northern region they appear to have a collective outward motion, and there is a grouping in the local neighbourhood of S\,Mon. While most Class II sources have an outward motion in the northern region, the position and kinematic behaviour of the majority of Class II sources in the NGC\,2264-C/D region is consistent with the properties of the `J',`K', and `M' sub-clusters identified by \citet{kuhn_spatial_2014} using finite mixture models and kinematically characterised by \citet{2019ApJ...870...32K} with DR2 data (see Figure 14 and Table 4 therein).

\begin{table}
\caption{Statistics of the $S_2$ (Class 0/I objects), $S_3$ (Class II objects), and $S_4$ (Class III objects) samples. The table lists the percentage of objects found to be spatially clustered ($I_5>I_{sig}$), median ($\tilde{I_5}$), and maximum ($\max I_5$) index values for each sample.} \label{table_stats}             
\centering                                      
\begin{tabular}{c c c c}         
\hline\hline                      
Sample & $I_5>I_{sig}$ & $\tilde{I_5}$ & $\max I_5$  \\ 
\hline 
$S_2$ & $84.7\%$ & 6.6 &  12.2 \\ 
$S_3$ & $35.2\%$ & 1.6  & 8.6  \\ 
$S_4$ & $2.8\%$ &  1.0 & 4.0  \\ 
\end{tabular}
\end{table}

\section{Discussion}\label{sect_discuss}

We summarise the results of our analysis as follows. There is a difference in spatial behaviour as a function of class in NGC\,2264. The youngest, most deeply embedded Class 0/I sources are typically found in strong concentrations with both the general population - and other Class 0/I - sources. While the more evolved Class II and TD sources are also found in such concentrations, the intensity of the concentrations and fraction of the population found in these significantly decreases with increasing evolutionary stage. The trend extends to the Class III sources for which the vast majority are randomly distributed and only a few are found in relatively loose concentrations with the general populous and/or sources of a similar class. This is consistent with previous studies of the region which identified, through qualitative analysis, that Class II objects as being more widely distributed than Class I objects (\citealt{2009AJ....138.1116S}, \citealt{2012A&A...540A..83T}). 

The spatial patterns we find are echoed in the kinematic behaviour of Class II and III sources, which differ considerably. Within the star formation active NGC\,2264-C/D regions, Class III sources have predominantly random trajectories and no clear groupings. Objects at the edge of the cluster typically have a larger proper motion than their more central counterparts, which is expected from virial balance as they see a larger enclosed mass. In contrast, Class II sources in this region do not have fully randomised trajectories and demonstrate kinematic behaviour consistent with the known substructure in \citet{2019ApJ...870...32K}. Although both samples are expected to contain some unresolved binaries, the disparity in their kinematic behaviours suggests that the observed spatial behaviour is age-driven dynamical evolution rather than primordial, this agrees with the work of \citet{2018A&A...609A..10V} who determined Class III objects to be older than Class II objects and to may have undergone post-birth migration.  

With age-driven dynamical evolution sources in the northern region should be significantly less clustered than the southern region, because star formation began there (\citealt{2009AJ....138.1116S}, \citealt{2010AJ....140.2070S}, \citealt{2017MmSAI..88..848V}, \citealt{2017MNRAS.465.1889G}, \citealt{2018A&A...609A..10V}), that is sources have had more time to disperse. Indeed, sources in the north are significantly less clustered than those of the south; $6.9\%$ and $29\%$ have an index above the significance threshold in the north and south respectively. Moreover, there is a correlation between the tightness of clusterings and region, with spatailly clustered sources having median $I_{5}$ values of 2.6 (north) and 4.6 (south). The outward motion observed in the kinematics for Class II objects in the north suggests a population at a more advanced stage of dispersal than their southern counterparts. 

Interestingly, in the vicinity of the northern O-type binary, S\,Mon, Class III objects display atypical spatial behaviour. Both \citet{2009AJ....138.1116S} and  \citet{2017MmSAI..88..848V} reported a lack of objects with discs within $0.1^{o}$ of the massive star due to disc disruption caused by stellar feedback. While we found Class III objects in the local neighbourhood of S\,Mon to be significantly more self-clustered within $0.1^{o}$ of S\,Mon than the wider NGC\,2264 region, the disparity is more prominent within $0.05^{o}$. Here, Class III's exhibit spatial behaviour patterns comparable to that of Class II sources suggesting disc ablation is causing these objects to appear as more evolved sources. 

\section{Conclusions}\label{sect_summary}

We have characterised the dynamic and spatial distributions of YSOs in the young NGC\,2264 cluster. This was achieved through analysis of pre-existing membership catalogues with the new local indicator tool INDICATE and kinematic data from the second instalment of the Gaia catalogue. 

In agreement with previous studies, we found the spatial behaviour of objects with and without discs to be distinct, indicating that star formation has been occurring sequentially over a prolonged period. The tool INDICATE has allowed us, for the first time, to quantitatively

\begin{enumerate}
\item establish spatial criteria for a source to be considered truly ‘clustered’ or ‘dispersed’ in the region; \newline
\item establish that the proportion of clustered sources decreases with increasing evolutionary stage (Class 0/I, II, TD, III);\newline
\item measure the tightness of these clusterings and establish that this decreases with increasing evolutionary stage; \newline
\item establish that the older northern region has a smaller proportion of clustered sources than the younger southern star formation active region;\newline
\item measure the tightness of these clusterings across the two regions and establish that they are tighter in the south than the north of the cluster; \newline
\item establish that Class IIIs within the local neighbourhood of S\,Mon exhibit spatial clustering behaviours typical of Class II objects in NGC\,2264. \newline
\end{enumerate}

Combining our spatial analysis with kinematic data from Gaia DR2 we derive strong evidence that NGC\,2264 is dynamically evolving with stars forming in a centralised, tightly clustered environment, in which they remain for their earliest stage of development before forming part of the dispersed population of NGC\,2264.  The effect of stellar feedback from S\,Mon on neighbouring stars is significant, causing these objects to appear as more evolved sources through disc ablation within a radius of $0.1^{o}$ and particularly within $0.05^{o}$.

Thanks to the second data release of Gaia an unprecedented volume of high-precision dynamical data became available for a large number of young clusters. With additional releases planned over the next few years our understanding of star formation and the nature of structures/patterns in these regions is set to increase profoundly. An important consideration going forward therefore is how best to extract, analyse, and interpret these data to produce reliable, robust, and consistent results.  In particular, it is important that the community gives careful consideration to terms relating to spatial distribution patterns of sources in these regions, such as ‘clustered’ and ‘dispersed’, especially in the context of identifying comparative differences. Until now such terms have been frequently used in literature as qualitative descriptors, but when applied subjectively to interpret dynamical behaviours they are at best vague, and at worst could lead to over-interpretation of the data. Building a true picture of star formation history in clusters will therefore require dynamical analyses to be validated by rigorous spatial analysis, in which such terms are clearly, consistently, and quantitatively defined. In this work, we demonstrated with NGC\,2264 that the local indicator code INDICATE which quantifies the intensity, correlation, and distribution of stars, can perform this analysis. When combined with Gaia DR2 data can be used to robustly analyse the star formation histories of young clusters.

\begin{acknowledgements}

The Star Form Mapper project has received funding from the European Union’s Horizon 2020 research and innovation programme under grant agreement No 687528.

\end{acknowledgements}

\bibliographystyle{aa} 
\bibliography{refs} 

\begin{thebibliography}{28}
\expandafter\ifx\csname natexlab\endcsname\relax\def\natexlab#1{#1}\fi

\bibitem[{{Anselin}(1995)}]{Anselin1995LISApaper}
{Anselin}, L. 1995, Geographical Analysis, 27, 93

\bibitem[{{Baglin} {et~al.}(2006){Baglin}, {Auvergne}, {Boisnard}, {Lam-Trong},
  {Barge}, {Catala}, {Deleuil}, {Michel}, \& {Weiss}}]{2006cosp...36.3749B}
{Baglin}, A., {Auvergne}, M., {Boisnard}, L., {et~al.} 2006, in 36th COSPAR
  Scientific Assembly, Vol.~36, 3749

\bibitem[{{Bailer-Jones} {et~al.}(2018){Bailer-Jones}, {Rybizki}, {Fouesneau},
  {Mantelet}, \& {Andrae}}]{2018AJ....156...58B}
{Bailer-Jones}, C.~A.~L., {Rybizki}, J., {Fouesneau}, M., {Mantelet}, G., \&
  {Andrae}, R. 2018, \aj, 156, 58

\bibitem[{{Buckner} {et~al.}(2019){Buckner}, {Khorrami}, {Khalaj}, {Lumsden},
  {Joncour}, {Moraux}, {Clark}, {Oudmaijer}, {Blanco}, {de la Calle},
  {Herrera-Fernandez}, {Motte}, {Salgado}, \&
  {Valero-Mart{\'{\i}}n}}]{2019A&A...622A.184B}
{Buckner}, A.~S.~M., {Khorrami}, Z., {Khalaj}, P., {et~al.} 2019, \aap, 622,
  A184

\bibitem[{{Cantat-Gaudin} {et~al.}(2018){Cantat-Gaudin}, {Jordi}, {Vallenari},
  {Bragaglia}, {Balaguer-N{\'u}{\~n}ez}, {Soubiran}, {Bossini}, {Moitinho},
  {Castro-Ginard}, {Krone-Martins}, {Casamiquela}, {Sordo}, \&
  {Carrera}}]{2018A&A...618A..93C}
{Cantat-Gaudin}, T., {Jordi}, C., {Vallenari}, A., {et~al.} 2018, \aap, 618,
  A93

\bibitem[{{Cody} {et~al.}(2014){Cody}, {Stauffer}, {Baglin}, {Micela},
  {Rebull}, {Flaccomio}, {Morales-Calder{\'o}n}, {Aigrain}, {Bouvier},
  {Hillenbrand}, {Gutermuth}, {Song}, {Turner}, {Alencar}, {Zwintz},
  {Plavchan}, {Carpenter}, {Findeisen}, {Carey}, {Terebey}, {Hartmann},
  {Calvet}, {Teixeira}, {Vrba}, {Wolk}, {Covey}, {Poppenhaeger}, {G{\"u}nther},
  {Forbrich}, {Whitney}, {Affer}, {Herbst}, {Hora}, {Barrado}, {Holtzman},
  {Marchis}, {Wood}, {Medeiros Guimar{\~a}es}, {Lillo Box}, {Gillen},
  {McQuillan}, {Espaillat}, {Allen}, {D'Alessio}, \&
  {Favata}}]{2014AJ....147...82C}
{Cody}, A.~M., {Stauffer}, J., {Baglin}, A., {et~al.} 2014, \aj, 147, 82

\bibitem[{{Fazio} {et~al.}(2004){Fazio}, {Hora}, {Allen}, {Ashby}, {Barmby},
  {Deutsch}, {Huang}, {Kleiner}, {Marengo}, {Megeath}, {Melnick}, {Pahre},
  {Patten}, {Polizotti}, {Smith}, {Taylor}, {Wang}, {Willner}, {Hoffmann},
  {Pipher}, {Forrest}, {McMurty}, {McCreight}, {McKelvey}, {McMurray}, {Koch},
  {Moseley}, {Arendt}, {Mentzell}, {Marx}, {Losch}, {Mayman}, {Eichhorn},
  {Krebs}, {Jhabvala}, {Gezari}, {Fixsen}, {Flores}, {Shakoorzadeh}, {Jungo},
  {Hakun}, {Workman}, {Karpati}, {Kichak}, {Whitley}, {Mann}, {Tollestrup},
  {Eisenhardt}, {Stern}, {Gorjian}, {Bhattacharya}, {Carey}, {Nelson},
  {Glaccum}, {Lacy}, {Lowrance}, {Laine}, {Reach}, {Stauffer}, {Surace},
  {Wilson}, {Wright}, {Hoffman}, {Domingo}, \& {Cohen}}]{2004ApJS..154...10F}
{Fazio}, G.~G., {Hora}, J.~L., {Allen}, L.~E., {et~al.} 2004, \apjs, 154, 10

\bibitem[{Feigelson {et~al.}(2013)Feigelson, Townsley, Broos, Busk, Getman,
  King, Kuhn, Naylor, Povich, Baddeley, Bate, Indebetouw, Luhman, McCaughrean,
  Pittard, Pudritz, Sills, Song, \& Wadsley}]{feigelson_overview_2013}
Feigelson, E.~D., Townsley, L.~K., Broos, P.~S., {et~al.} 2013, Astrophysical
  $Journal$s, 209, 26

\bibitem[{{Gaia Collaboration} {et~al.}(2018){Gaia Collaboration}, {Brown},
  {Vallenari}, {Prusti}, {de Bruijne}, {Babusiaux}, {Bailer-Jones}, {Biermann},
  {Evans}, {Eyer}, \& et~al.}]{2018A&A...616A...1G}
{Gaia Collaboration}, {Brown}, A.~G.~A., {Vallenari}, A., {et~al.} 2018, \aap,
  616, A1

\bibitem[{{Gonz{\'a}lez} \& {Alfaro}(2017)}]{2017MNRAS.465.1889G}
{Gonz{\'a}lez}, M. \& {Alfaro}, E.~J. 2017, \mnras, 465, 1889

\bibitem[{{Gutermuth} {et~al.}(2009){Gutermuth}, {Megeath}, {Myers}, {Allen},
  {Pipher}, \& {Fazio}}]{2009ApJS..184...18G}
{Gutermuth}, R.~A., {Megeath}, S.~T., {Myers}, P.~C., {et~al.} 2009, The
  Astrophysical Journal Supplement Series, 184, 18

\bibitem[{Kuhn {et~al.}(2014)Kuhn, Feigelson, Getman, Baddeley, Broos, Sills,
  Bate, Povich, Luhman, Busk, Naylor, \& King}]{kuhn_spatial_2014}
Kuhn, M.~A., Feigelson, E.~D., Getman, K.~V., {et~al.} 2014, Astrophysical
  $Journal$, 787, 107

\bibitem[{{Kuhn} {et~al.}(2019){Kuhn}, {Hillenbrand}, {Sills}, {Feigelson}, \&
  {Getman}}]{2019ApJ...870...32K}
{Kuhn}, M.~A., {Hillenbrand}, L.~A., {Sills}, A., {Feigelson}, E.~D., \&
  {Getman}, K.~V. 2019, \apj, 870, 32

\bibitem[{{Lawrence} {et~al.}(2007){Lawrence}, {Warren}, {Almaini}, {Edge},
  {Hambly}, {Jameson}, {Lucas}, {Casali}, {Adamson}, {Dye}, {Emerson},
  {Foucaud}, {Hewett}, {Hirst}, {Hodgkin}, {Irwin}, {Lodieu}, {McMahon},
  {Simpson}, {Smail}, {Mortlock}, \& {Folger}}]{2007MNRAS.379.1599L}
{Lawrence}, A., {Warren}, S.~J., {Almaini}, O., {et~al.} 2007, \mnras, 379,
  1599

\bibitem[{{Lindegren} {et~al.}(2018){Lindegren}, {Hern{\'a}ndez}, {Bombrun},
  {Klioner}, {Bastian}, {Ramos-Lerate}, {de Torres}, {Steidelm{\"u}ller},
  {Stephenson}, {Hobbs}, {Lammers}, {Biermann}, {Geyer}, {Hilger}, {Michalik},
  {Stampa}, {McMillan}, {Casta{\~n}eda}, {Clotet}, {Comoretto}, {Davidson},
  {Fabricius}, {Gracia}, {Hambly}, {Hutton}, {Mora}, {Portell}, {van Leeuwen},
  {Abbas}, {Abreu}, {Altmann}, {Andrei}, {Anglada}, {Balaguer-N{\'u}{\~n}ez},
  {Barache}, {Becciani}, {Bertone}, {Bianchi}, {Bouquillon}, {Bourda},
  {Br{\"u}semeister}, {Bucciarelli}, {Busonero}, {Buzzi}, {Cancelliere},
  {Carlucci}, {Charlot}, {Cheek}, {Crosta}, {Crowley}, {de Bruijne}, {de
  Felice}, {Drimmel}, {Esquej}, {Fienga}, {Fraile}, {Gai}, {Garralda},
  {Gonz{\'a}lez-Vidal}, {Guerra}, {Hauser}, {Hofmann}, {Holl}, {Jordan},
  {Lattanzi}, {Lenhardt}, {Liao}, {Licata}, {Lister}, {L{\"o}ffler},
  {Marchant}, {Martin-Fleitas}, {Messineo}, {Mignard}, {Morbidelli}, {Poggio},
  {Riva}, {Rowell}, {Salguero}, {Sarasso}, {Sciacca}, {Siddiqui}, {Smart},
  {Spagna}, {Steele}, {Taris}, {Torra}, {van Elteren}, {van Reeven}, \&
  {Vecchiato}}]{2018A&A...616A...2L}
{Lindegren}, L., {Hern{\'a}ndez}, J., {Bombrun}, A., {et~al.} 2018, \aap, 616,
  A2

\bibitem[{{Mayne} \& {Naylor}(2008)}]{2008MNRAS.386..261M}
{Mayne}, N.~J. \& {Naylor}, T. 2008, \mnras, 386, 261

\bibitem[{{Peretto} {et~al.}(2006){Peretto}, {Andr{\'e}}, \&
  {Belloche}}]{2006A&A...445..979P}
{Peretto}, N., {Andr{\'e}}, P., \& {Belloche}, A. 2006, \aap, 445, 979

\bibitem[{{Povich} {et~al.}(2013){Povich}, {Kuhn}, {Getman}, {Busk},
  {Feigelson}, {Broos}, {Townsley}, {King}, \& {Naylor}}]{2013ApJS..209...31P}
{Povich}, M.~S., {Kuhn}, M.~A., {Getman}, K.~V., {et~al.} 2013, \apjs, 209, 31

\bibitem[{{Rapson} {et~al.}(2014){Rapson}, {Pipher}, {Gutermuth}, {Megeath},
  {Allen}, {Myers}, \& {Allen}}]{2014ApJ...794..124R}
{Rapson}, V.~A., {Pipher}, J.~L., {Gutermuth}, R.~A., {et~al.} 2014, \apj, 794,
  124

\bibitem[{{Skrutskie} {et~al.}(2006){Skrutskie}, {Cutri}, {Stiening},
  {Weinberg}, {Schneider}, {Carpenter}, {Beichman}, {Capps}, {Chester},
  {Elias}, {Huchra}, {Liebert}, {Lonsdale}, {Monet}, {Price}, {Seitzer},
  {Jarrett}, {Kirkpatrick}, {Gizis}, {Howard}, {Evans}, {Fowler}, {Fullmer},
  {Hurt}, {Light}, {Kopan}, {Marsh}, {McCallon}, {Tam}, {Van Dyk}, \&
  {Wheelock}}]{2006AJ....131.1163S}
{Skrutskie}, M.~F., {Cutri}, R.~M., {Stiening}, R., {et~al.} 2006, \aj, 131,
  1163

\bibitem[{{Sung} \& {Bessell}(2010)}]{2010AJ....140.2070S}
{Sung}, H. \& {Bessell}, M.~S. 2010, \aj, 140, 2070

\bibitem[{{Sung} {et~al.}(2009){Sung}, {Stauffer}, \&
  {Bessell}}]{2009AJ....138.1116S}
{Sung}, H., {Stauffer}, J.~R., \& {Bessell}, M.~S. 2009, Astronomical
  $Journal$, 138, 1116

\bibitem[{{Teixeira} {et~al.}(2012){Teixeira}, {Lada}, {Marengo}, \&
  {Lada}}]{2012A&A...540A..83T}
{Teixeira}, P.~S., {Lada}, C.~J., {Marengo}, M., \& {Lada}, E.~A. 2012, \aap,
  540, A83

\bibitem[{{Teixeira} {et~al.}(2007){Teixeira}, {Zapata}, \&
  {Lada}}]{2007ApJ...667L.179T}
{Teixeira}, P.~S., {Zapata}, L.~A., \& {Lada}, C.~J. 2007, \apjl, 667, L179

\bibitem[{{van Leeuwen}(2009)}]{vanLeeuwen_2009}
{van Leeuwen}, F. 2009, \aap, 497, 209

\bibitem[{{Venuti} {et~al.}(2017){Venuti}, {Prisinzano}, {Sacco}, {Flaccomio},
  {Bonito}, {Damiani}, {Micela}, {Guarcello}, {GES Collaboration}, \& {CSI 2264
  Collaboration}}]{2017MmSAI..88..848V}
{Venuti}, L., {Prisinzano}, L., {Sacco}, G., {et~al.} 2017, \memsai, 88, 848

\bibitem[{{Venuti} {et~al.}(2018){Venuti}, {Prisinzano}, {Sacco}, {Flaccomio},
  {Bonito}, {Damiani}, {Micela}, {Guarcello}, {Randich}, {Stauffer}, {Cody},
  {Jeffries}, {Alencar}, {Alfaro}, {Lanzafame}, {Pancino}, {Bayo}, {Carraro},
  {Costado}, {Frasca}, {Jofr{\'e}}, {Morbidelli}, {Sousa}, \&
  {Zaggia}}]{2018A&A...609A..10V}
{Venuti}, L., {Prisinzano}, L., {Sacco}, G.~G., {et~al.} 2018, \aap, 609, A10

\bibitem[{{Weisskopf} {et~al.}(2000){Weisskopf}, {Tananbaum}, {Van Speybroeck},
  \& {O'Dell}}]{2000SPIE.4012....2W}
{Weisskopf}, M.~C., {Tananbaum}, H.~D., {Van Speybroeck}, L.~P., \& {O'Dell},
  S.~L. 2000, Society of Photo-Optical Instrumentation Engineers (SPIE)
  Conference Series, Vol. 4012, {Chandra X-ray Observatory (CXO): overview},
  ed. J.~E. {Truemper} \& B.~{Aschenbach}, 2--16

\end{thebibliography}

\clearpage

\begin{appendix}

\section{Proper motion perspective corrections}\label{append_gaia_pm}

We correct for the perspective contraction of NGC\,2264 (caused by radial motions of members) for each source $i$ using Eq.13 of \citet{vanLeeuwen_2009} as follows:

\begin{equation}\label{eq_pmracor}
\\ \mu_{\alpha*,\,i}^\mathrm{cor}=\Delta\,\alpha_{i}\left(\mu_{\delta_{0}}\sin\,\delta_{0}-\frac{V_\mathrm{r0}\omega_{0}}{\kappa}\cos\,\delta_{0}\right)
\end{equation}

\begin{equation}\label{eq_pmdeccor}
\\ \mu_{\delta,\,i}^\mathrm{cor}=-\Delta\,\alpha_{i}\,\mu_{\alpha^{*}_{0}}\sin\,\delta_{0}-\frac{\Delta\,\delta_{i}V_\mathrm{r0}\omega_{0}}{\kappa}
\end{equation}

where $\mu_{\alpha*,\,i}^\mathrm{cor}$ and $\mu_{\delta,\,i}^\mathrm{cor}$ are the corrected components of proper motion in right ascension and declination respectively. Table\,\ref{table_cons} lists the values of the central coordinates of the cluster($\alpha_{0},\delta_{0}$), distance unit conversion factor ($\kappa$); mean proper motion ($\mu_{\alpha^{*}_{0}},\mu_{\delta_{0}}$), parallax ($\omega_{0}$) and radial velocity ($V_\mathrm{r0}$) used.

For each source we subtract the perspective correction and mean proper motion of the sample to gain the corrected internal proper motion:

\begin{equation}
\\ \mu_{\alpha*,\,i}^\mathrm{final}=\mu_{\alpha*,\,i}^\mathrm{DR2} -\mu_{\alpha*,\,i}^\mathrm{cor} - \mu_{\alpha^{*}_{0}}
\end{equation}
\begin{equation}
\\ \mu_{\delta,\,i}^\mathrm{final}=\mu_{\delta,\,i}^\mathrm{DR2} - \mu_{\delta,\,i}^\mathrm{cor} -\mu_{\delta_{0}}
\end{equation}

where ($\mu_{\alpha*,\,i}^\mathrm{final}$, $\mu_{\delta,\,i}^\mathrm{final}$) are the corrected, and ($\mu_{\alpha*,\,i}^\mathrm{DR2}, \mu_{\delta,\,i}^\mathrm{DR2}$) the Gaia DR2, components of proper motion for source $i$ in right ascension and declination respectively.

\begin{table}
\caption{Values and sources of constants used in Eqs.\,\ref{eq_pmracor} and \ref{eq_pmdeccor}. \label{table_cons}}             
\label{table_example}     
\centering                                     
\begin{tabular}{c c c }         
\hline\hline                        
Constant & Value & Source  \\ 
\hline 
$\alpha_{0}$ & 100.241 deg & \citet{kuhn_spatial_2014}\\ 
$\delta_{0}$ & 9.680 deg & \citet{kuhn_spatial_2014}\\
$\omega_{0}$ & 1.363$\pm$0.003 mas  &  This Paper \\  
$\mu_{\alpha_{0}}$ & -1.817$\pm$0.005 mas\,yr$^{-1}$ &  This Paper  \\ 
$\mu_{\delta_{0}}$ & -3.919$\pm$0.004 mas\,yr$^{-1}$ &  This Paper  \\ 
$V_\mathrm{r0}$ & 16.6$\pm$1.0 km\,s$^{-1}$  &  This Paper \\ 
$\kappa$ & 4.74 & \citet{vanLeeuwen_2009} \\ 
\end{tabular}
\end{table}

\clearpage
\onecolumn
\section{Inconsistent catalogue classifications}
\begin{longtable}{c l c c }
\caption{List of cross-matches between the R14 and K14 catalogues with inconsistent classifications. \label{appen_duplic}} \\      
\hline \hline 
[R14] ID & [K14] MCPM & [R14] Class & [K14] Class \\ 
\hline
\endfirsthead

\caption{continued.}\\
\hline\hline
[R14] ID & [K14] MCPM & [R14] Class & [K14] Class \\ 

\hline 
\endhead
\hline
\endfoot
\endlastfoot
22250 & 064119.40+092146.7 & III/F & 0/I\\ 
22730 & 064112.30+092224.2 & III/F & 0/I\\ 
24628 & 064111.30+092459.3 & 0/I & Amb\\ 
24672 & 064106.22+092503.6 & II & Amb\\ 
24792 & 064034.16+092512.7 & III/F & Amb\\ 
24857 & 064034.33+092517.0 & II & Amb\\ 
25220 & 064109.64+092545.4 & 0/I & II/III\\ 
25346 & 064114.87+092555.2 & II & 0/I\\ 
25606 & 064112.86+092614.9 & II & 0/I\\ 
25722 & 064052.94+092625.7 & II & 0/I\\ 
26143 & 064106.43+092658.6 & 0/I & II/III\\ 
26299 & 064116.18+092710.6 & 0/I & Amb\\ 
26560 & 064107.12+092728.9 & II & Amb\\ 
26657 & 064113.41+092736.2 & II & Amb\\ 
27066 & 064117.51+092806.3 & II & Amb\\ 
27157 & 064117.55+092813.2 & 0/I & Amb\\ 
27411 & 064120.09+092834.7 & 0/I & Amb\\ 
27413 & 064100.28+092833.9 & II & Amb\\ 
27526 & 064059.68+092843.8 & II & Amb\\ 
27527 & 064052.72+092843.7 & II & Amb\\ 
27535 & 064120.70+092845.4 & II & Amb\\ 
27536 & 064108.49+092844.6 & 0/I & Amb\\ 
27574 & 064037.22+092847.0 & AGN & II/III\\ 
27751 & 064117.92+092901.1 & II & Amb\\ 
27919 & 064052.09+092913.8 & II & Amb\\ 
28069 & 064106.90+092924.0 & II & Amb\\ 
28083 & 064109.53+092925.3 & II & Amb\\ 
28100 & 064038.33+092925.5 & III/F & Amb\\ 
28157 & 064118.30+092932.4 & III/F & 0/I\\ 
28343 & 064108.92+092944.9 & 0/I & II/III\\ 
28427 & 064059.49+092951.6 & II & Amb\\ 
28516 & 064117.63+092958.8 & II & 0/I\\ 
28583 & 064056.66+093002.8 & II & 0/I\\ 
28598 & 064108.19+093003.8 & II & 0/I\\ 
28668 & 064108.17+093007.8 & 0/I & Amb\\ 
28676 & 064109.08+093009.0 & II & 0/I\\ 
28714 & 064113.31+093012.0 & II & 0/I\\ 
28870 & 064116.80+093022.4 & II & Amb\\ 
28883 & 064108.27+093022.8 & II & 0/I\\ 
28884 & 064053.39+093022.5 & PAH & II/III\\ 
28904 & 064113.42+093023.6 & II & 0/I\\ 
28938 & 064109.30+093025.6 & 0/I & II/III\\ 
28972 & 064107.61+093029.2 & II & 0/I\\ 
29054 & 064043.40+093034.2 & II & 0/I\\ 
29095 & 064115.88+093037.3 & II & 0/I\\ 
29357 & 064058.81+093057.1 & II & 0/I\\ 
29479 & 064108.56+093105.8 & II & Amb\\ 
29495 & 064106.57+093106.6 & II & Amb\\ 
29524 & 064109.01+093108.7 & II & 0/I\\ 
29716 & 064115.30+093122.1 & II & 0/I\\ 
30104 & 064113.28+093150.3 & II & Amb\\ 
30699 & 064123.30+093230.1 & AGN & 0/I\\ 
31083 & 064056.99+093301.3 & 0/I & Amb\\ 
31393 & 064104.23+093323.6 & 0/I & Amb\\ 
31415 & 064059.26+093325.0 & II & Amb\\ 
31417 & 064053.63+093324.7 & TD & 0/I\\ 
31472 & 064106.70+093330.0 & 0/I & Amb\\ 
31509 & 064104.23+093332.0 & AGN & Amb\\ 
31521 & 064113.42+093332.9 & III/F & 0/I\\ 
31532 & 064059.36+093333.3 & II & 0/I\\ 
31551 & 064042.77+093334.9 & III/F & Amb\\ 
31661 & 064104.47+093343.8 & II & Amb\\ 
31750 & 064106.31+093350.0 & 0/I & Amb\\ 
31815 & 064105.61+093355.0 & II & 0/I\\ 
31851 & 064106.65+093357.6 & 0/I & Amb\\ 
31881 & 064054.40+093358.9 & AGN & Amb\\ 
31980 & 064105.72+093406.4 & 0/I & Amb\\ 
32004 & 064110.92+093408.2 & II & 0/I\\ 
32081 & 064111.06+093412.3 & II & Amb\\ 
32103 & 064114.76+093413.6 & II & Amb\\ 
32104 & 064105.38+093413.2 & 0/I & Amb\\ 
32145 & 064110.42+093418.8 & 0/I & Amb\\ 
32183 & 064106.66+093420.7 & AGN & II/III\\ 
32330 & 064052.39+093431.4 & II & 0/I\\ 
32383 & 064101.82+093434.1 & 0/I & Amb\\ 
32474 & 064030.86+093440.5 & II & Amb\\ 
32489 & 064125.62+093442.9 & II & Amb\\ 
32531 & 064106.73+093445.9 & II & Amb\\ 
32533 & 064039.34+093445.5 & II & Amb\\ 
32540 & 064107.98+093446.8 & II & Amb\\ 
32625 & 064101.33+093452.6 & II & Amb\\ 
32641 & 064050.30+093453.7 & 0/I & II/III\\ 
32647 & 064107.39+093454.9 & II & 0/I\\ 
32715 & 064104.26+093459.5 & 0/I & II/III\\ 
32728 & 064059.97+093500.8 & II & 0/I\\ 
32737 & 064040.51+093501.1 & II & 0/I\\ 
32898 & 064111.84+093514.4 & 0/I & Amb\\ 
33104 & 064105.77+093529.5 & II & 0/I\\ 
33124 & 064111.83+093531.4 & II & 0/I\\ 
33150 & 064104.29+093533.2 & II & 0/I\\ 
33165 & 064102.81+093534.3 & II & Amb\\ 
33189 & 064114.19+093535.5 & AGN & II/III\\ 
33190 & 064105.23+093535.7 & 0/I & Amb\\ 
33191 & 064050.40+093535.9 & II & Amb\\ 
33228 & 064104.00+093538.3 & II & 0/I\\ 
33306 & 064105.60+093544.4 & 0/I & Amb\\ 
33320 & 064105.91+093545.0 & III/F & 0/I\\ 
33321 & 064103.13+093544.9 & II & 0/I\\ 
33405 & 064059.29+093552.3 & II & 0/I\\ 
33485 & 064100.28+093558.9 & II & 0/I\\ 
33486 & 064059.75+093559.1 & II & Amb\\ 
33523 & 064058.95+093601.0 & II & Amb\\ 
33544 & 064108.65+093603.2 & 0/I & II/III\\ 
33568 & 064103.61+093604.4 & II & Amb\\ 
33582 & 064055.78+093606.0 & II & 0/I\\ 
33633 & 064100.64+093610.0 & II & Amb\\ 
33634 & 064059.52+093610.4 & 0/I & Amb\\ 
33635 & 064051.85+093609.9 & AGN & II/III\\ 
33685 & 064058.00+093614.5 & 0/I & Amb\\ 
33712 & 064114.11+093616.4 & AGN & II/III\\ 
33713 & 064102.79+093616.0 & II & Amb\\ 
33745 & 064104.61+093618.1 & 0/I & Amb\\ 
33863 & 064057.39+093628.2 & AGN & Amb\\ 
33896 & 064105.37+093630.6 & 0/I & II/III\\ 
33897 & 064100.24+093631.1 & II & Amb\\ 
33898 & 064056.17+093630.9 & II & 0/I\\ 
33951 & 064059.82+093633.3 & II & 0/I\\ 
34032 & 064058.61+093639.3 & II & 0/I\\ 
34041 & 064102.58+093640.1 & II & Amb\\ 
34087 & 064104.43+093643.3 & II & Amb\\ 
34226 & 064058.09+093653.3 & II & Amb\\ 
34259 & 064108.19+093656.0 & II & Amb\\ 
34271 & 064059.62+093657.5 & III/F & Amb\\ 
34481 & 064049.18+093714.3 & II & Amb\\ 
34624 & 064101.62+093728.5 & II & 0/I\\ 
34679 & 064123.14+093733.9 & II & 0/I\\ 
34708 & 064049.12+093736.3 & II & Amb\\ 
34800 & 064111.90+093743.8 & II & 0/I\\ 
34944 & 064108.89+093754.6 & 0/I & Amb\\ 
34974 & 064058.33+093756.7 & II & Amb\\ 
34981 & 064115.20+093757.6 & II & Amb\\ 
35389 & 064104.56+093830.8 & II & Amb\\ 
35864 & 064059.54+093906.3 & II & Amb\\ 
35946 & 064105.98+093914.0 & II & 0/I\\ 
36165 & 064056.24+093932.6 & II & 0/I\\ 
36393 & 064102.17+093951.4 & 0/I & Amb\\ 
37159 & 064020.63+094049.9 & II & 0/I\\ 
38126 & 064029.78+094221.1 & II & Amb\\ 
38366 & 064101.73+094242.9 & II & Amb\\ 
39195 & 064125.62+094403.3 & II & 0/I\\ 
39518 & 064104.83+094433.2 & II & 0/I\\ 
39622 & 064041.02+094442.4 & II & 0/I\\ 
40462 & 064042.26+094607.9 & III/F & 0/I\\ 
40472 & 064024.77+094607.9 & II & 0/I\\ 
40810 & 064114.84+094646.7 & III/F & Amb\\ 
40929 & 064124.03+094700.7 & II & 0/I\\ 
40972 & 064059.90+094704.5 & II & 0/I\\ 
40974 & 064053.62+094704.3 & II & Amb\\ 
41308 & 064037.05+094736.0 & II & 0/I\\ 
41309 & 064029.41+094736.9 & II & Amb\\ 
41615 & 064039.36+094806.2 & II & 0/I\\ 
41982 & 064054.13+094843.4 & II & Amb\\ 
42191 & 064059.68+094904.6 & II & 0/I\\ 
42362 & 064054.26+094920.3 & II & Amb\\ 
42384 & 064105.06+094922.7 & II & Amb\\ 
42494 & 064031.10+094931.9 & II & Amb\\ 
42532 & 064036.09+094935.2 & II & 0/I\\ 
42533 & 064032.03+094935.3 & II & Amb\\ 
42557 & 064030.07+094937.6 & 0/I & II/III\\ 
42718 & 064033.11+094954.7 & II & Amb\\ 
42824 & 064033.26+095006.2 & II & 0/I\\ 
43021 & 064028.55+095028.8 & II & 0/I\\ 
43114 & 064051.14+095037.9 & II & Amb\\ 
43155 & 064029.85+095043.4 & AGN & 0/I\\ 
43236 & 064027.58+095051.6 & II & Amb\\ 
43291 & 064025.89+095057.3 & II & 0/I\\ 
43393 & 064057.84+095108.9 & II & Amb\\ 
43680 & 064046.24+095140.0 & II & 0/I\\ 
43705 & 064048.90+095144.4 & II & 0/I\\ 
43706 & 064041.84+095144.5 & II & Amb\\ 
43928 & 064100.80+095207.5 & II & Amb\\ 
44176 & 064112.57+095231.1 & II & Amb\\ 
44270 & 064046.95+095240.5 & TD & 0/I\\ 
44352 & 064116.42+095249.6 & AGN & II/III\\ 
44395 & 064049.37+095253.8 & II & Amb\\ 
44551 & 064037.21+095310.3 & II & 0/I\\ 
44572 & 064054.88+095312.3 & II & 0/I\\ 
45085 & 064016.11+095407.2 & II & 0/I\\ 
45264 & 064036.36+095427.0 & II & Amb\\ 
45295 & 064025.51+095432.3 & II & Amb\\ 
45382 & 064104.56+095443.8 & II & Amb\\ 
45447 & 064117.38+095451.1 & AGN & Amb\\ 
45490 & 064023.42+095455.5 & II & Amb\\ 
45751 & 064023.73+095523.8 & II & Amb\\ 
46986 & 064110.70+095742.4 & II & 0/I\\ 
47241 & 064057.57+095812.7 & II & 0/I\\ 
47933 & 064059.46+095945.4 & II & Amb\\  
\end{longtable}
\clearpage
\twocolumn

\end{appendix}

\end{document}